\newcommand{\vece}{\mbox{\boldmath$e$}}

\newcommand{\vecj}{\mbox{\boldmath$j$}}

\newcommand{\vecp}{\mbox{\boldmath$p$}}

\newcommand{\vecr}{\mbox{\boldmath$r$}}

\newcommand{\vecv}{\mbox{\boldmath$v$}}

\newcommand{\vecA}{\mbox{\boldmath$A$}}
\newcommand{\vecB}{\mbox{\boldmath$B$}}

\newcommand{\vecE}{\mbox{\boldmath$E$}}

\newcommand{\vecH}{\mbox{\boldmath$H$}}

\newcommand{\vecnl}{\mbox{\boldmath$0$}}

\newcommand{\dfd}{{\rm d}}

\newcommand{\half}{\frac{1}{2}}

\newcommand{\ie}{{\em i.e.}}
\newcommand{\eg}{{\em e.g.}}

\documentclass[prb,showpacs]{revtex4}% Physical Review B

\usepackage{graphicx}% Include figure files
\usepackage{dcolumn}% Align table columns on decimal point
\usepackage{bm}% bold math

\begin{document}

%\preprint{APS/123-QED}

\title{Meissner effect, diamagnetism, and classical physics -- a  review}% Force line breaks with \\

\author{Hanno Ess{\'e}n}
% \altaffiliation[Also at ]{}%Lines break automatically or can be forced with \\
\affiliation{Department of Mechanics, Royal Institute of Technology (KTH)\\ Stockholm SE-100 44, Sweden
}%
\author{Miguel C. N. Fiolhais}%
\affiliation{LIP-Coimbra, Department of Physics\\ University of Coimbra, Coimbra 3004-516, Portugal
}%

\date{October 27, 2011}% It is always \today, today,
             %  but any date may be explicitly specified

\begin{abstract}
We review the literature on what classical physics has to say about the Meissner effect and the
London equations. We discuss the relevance of the Bohr-van Leeuwen theorem for the  perfect
diamagnetism of superconductors and conclude that the theorem is based on invalid assumptions. We
also point out results in the literature that show how magnetic flux expulsion from a sample cooled
to superconductivity can be understood as an approach to the magnetostatic energy minimum. These
results have been published several times but many textbooks on magnetism still claim that there is
no classical diamagnetism, and virtually all books on superconductivity repeat Meissner's 1933
statement that flux expulsion has no classical explanation.\\ \\
The following article has been accepted for publication in American Journal of Physics. When it is published, it will be found at http://scitation.aip.org/ajp/.
\end{abstract}

% PACS, the Physics and Astronomy
                             % Classification Scheme.
%\keywords{Suggested keywords}%Use showkeys class option if keyword
                              %display desired
\maketitle

\section{Introduction}
It is now a century since superconductivity was discovered by Kammerlingh-Onnes in Leiden in 1911.
From the beginning, there was considerable interest from theoretical physicists. Unfortunately
progress has been slow and one can safely say that the phenomenon is still not completely understood, at least not in a fundamental reductionist sense. It is therefore important that the
things that {\em can} be understood are correctly presented in textbooks. To the contrary, textbooks often repeat two myths which have become so ingrained in the minds of
physicists that they hamper progress. These are:
\begin{itemize}
    \item There is no classical diamagnetism (Bohr 1911,\cite{bohr} van Leeuwen 1919\cite{van_leeuwen}).
    \item There is no classical explanation of flux expulsion from a superconductor (Meissner
        1933\cite{SCmeissner}).
\end{itemize}
Although both statements have been disproved multiple times in the scientific literature, these myths continue to be spread. We hope this review will improve the situation.

Meissner found it natural that a weak magnetic field could not penetrate a Type-I superconductor. The fact that this is already in conflict with
classical physics according to the Bohr-van Leeuwen theorem is rarely mentioned. On the other hand, Meissner found it remarkable that a normal metal with a magnetic field inside will expel this field
when cooled to superconductivity. This was claimed to have no classical explanation, and
consequently superconductors were not considered perfect conductors.

We will not present any new results in this paper.  Instead, we will review some of the strong evidence in the
archival literature demonstrating that the above two statements are false. We begin with the Bohr-van
Leeuwen theorem and then turn to the classical explanation of flux expulsion. After discussing the
arguments for the traditional point of view, we will demonstrate why they are wrong.

\section{The Bohr-van Leeuwen theorem and classical diamagnetism}
The Hamiltonian $H$, for a system of charged particles interacting via a (scalar) potential energy $U$
is
\begin{equation}\label{eq.hamiltonian.no.field}
H(\vecr_k,\vecp_k) =  \sum_{j=1}^N \frac{\vecp_j^2}{2m_j} + U(\vecr_k) .
\end{equation}
The particles have masses $m_j$, position vectors $\vecr_j$, and momenta $\vecp_j$. The presence of
an external magnetic field with vector potential $\vecA(\vecr)$ alters the Hamiltonian to
\begin{equation}\label{eq.hamiltonian.ex.field}
H(\vecr_k,\vecp_k) =  \sum_{j=1}^N \frac{\left[\vecp_j -\frac{e_j}{c}\vecA(\vecr_j)\right]^2 }{2m_j}
+ U(\vecr_k),
\end{equation}
where $e_j$ are the charges of masses $m_j$. Using this equation one can show that
statistical mechanics predicts that the energy---the thermal average of
the Hamiltonian---does not depend on the external field. Hence the system exhibits neither a
paramagnetic nor a diamagnetic response.

In his 1911 doctoral dissertation, Niels Bohr \cite{bohr} used the above equation and concluded there is no magnetic response of a metal according to classical physics. In 1919 Hendrika
Johanna van Leeuwen independently came to the same conclusion in her Leiden thesis. When her
work was published in 1921, Miss van Leeuwen \cite{van_leeuwen} noted that similar conclusions had
been reached by Bohr.

Many books on magnetism refer to the above results as the Bohr-van Leeuwen theorem, or simply the van
Leeuwen theorem. The theorem is often summarized as stating there is no classical magnetism. Since
this is obvious in the case of spin or atomic angular momenta---the quantum phenomena responsible
for paramagnetism---the more interesting conclusion is that classical statistical mechanics and
electromagnetism cannot explain diamagnetism. A very thorough treatment of the Bohr-van Leeuwen theorem can be found in Van Vleck.\cite{BKvanvleck} Other books, such as Mohn,\cite{BKmohn} Getzlaff,\cite{BKgetzlaff} Aharoni,\cite{BKaharoni} and  L{\'e}vy,\cite{BKlevy} mention the theorem more or less briefly. Weaknesses in classical derivations of diamagnetism in modern textbooks have been pointed out by O'Dell and Zia.\cite{odell&zia} A discussion of the theorem can also be found in the Feynman lectures\cite{BKfeynman&al2} (vol. II section 34-6), which points out that a constant external field will not do any work on a system of charges. Therefore, the energy of this system cannot depend on the external field. In addition, the fact that the diamagnetic response of most materials is very small lends empirical support to the theorem.

\subsection{The inadequacy of the basic assumptions}
If we accept the popular version of the Bohr-van Leeuwen theorem as true, then one must conclude
that the perfect diamagnetism of superconductors cannot have a classical explanation. A study of the
proof of the theorem, however, reveals that it is only valid under assumptions that do not
necessarily hold. The assumed Hamiltonian of Eq.~(\ref{eq.hamiltonian.ex.field}) only includes the
vector potential of the external magnetic field. But it has been known since 1920 that the best
Hamiltonian for a system of classical charged particles is the Darwin Hamiltonian.\cite{darwin} In
this Hamiltonian one takes into account the {\em internal} magnetic fields generated by the moving charged
particles of the system itself, plus any corresponding interactions.

The simplest way to see that the total energy of a system of charged particles depends on the
external field is to note that the total energy includes a magnetic energy\cite{Footnote12}
\begin{equation}\label{eq.mag.energy.1.2}
E_B = \frac{1}{8\pi} \int \vecB^2(\vecr)\, \dfd V  =  \frac{1}{8\pi}\int  (\vecB_{\rm e} +
\vecB_{\rm i})^2 \,\dfd V ,
\end{equation}
where $\vecB_{\rm e}$ is the external magnetic field and $\vecB_{\rm i}$ is the internal field. Using the
Biot-Savart law this field is given, to first order in $v/c$, by
\begin{equation}\label{eq.int.mag.field}
\vecB_{\rm i}(\vecr) =\sum_{j=1}^N \vecB_j (\vecr) = \sum_{j=1}^N \frac{e_j}{c} \frac{\vecv_j \times
(\vecr - \vecr_j)}{|\vecr - \vecr_j|^3}.
\end{equation}
Equation (\ref{eq.mag.energy.1.2}) makes it obvious that to minimize the energy the internal field
will be, as much as possible, equal in magnitude and opposite in direction to the external
field. This gives diamagnetism.

Charles Galton Darwin was first to derive an approximate Lagrangian for a system of charged particles (neglecting radiation) that is correct to order $(v/c)^2$.\cite{darwin,BKlandau2,breitenberger,kennedy,essen96} In a time independent external magnetic field, there is then a conserved Darwin energy given by
\begin{equation}\label{eq.energy.darw}
E_D = \sum_{j=1}^N \left  [\frac{m_j}{2} \vecv_j^2 + \frac{e_j}{c} \vecv_j \cdot \left(\half
\vecA_{\rm i}(\vecr_j;\vecr_k,\vecv_k) + \vecA_{\rm e}(\vecr_j) \right)\right] + U(\vecr_k)+E_{\rm
e}.
\end{equation}
In this equation, $\vecv_j$ are velocity vectors, $\vecA_{\rm e}$ is the external vector potential,
$E_{\rm e}$ is the (constant) energy of the external magnetic field, and $\vecA_{\rm i}$ is the
internal vector potential given by
\begin{equation}\label{eq.darw.vec.pot}
\vecA_{\rm i}(\vecr_j;\vecr_k,\vecv_k)=\sum_{k\neq j}^N \frac{e_k}{r_{kj}}\frac{\vecv_k +
(\vecv_k\cdot \vece_{kj})\vece_{kj}}{2c},
\end{equation}
where $r_{kj}=|\vecr_j - \vecr_k|$ and $\vece_{kj}=(\vecr_j - \vecr_k)/r_{kj}$.  Although there is
a corresponding Hamiltonian, it cannot be written in closed form. When the Darwin magnetic
interactions are taken into account, the Bohr-van Leeuwen theorem is no longer valid because the
magnetic fields of the moving charges will contribute to the total magnetic energy. The fact that
this invalidates the Bohr-van Leeuwen theorem for superconductors was stated explicitly by
Pfleiderer\cite{SCpfleiderer} in a letter to Nature in 1966. A more recent discussion of classical
diamagnetism and the Darwin Hamiltonian is given by Ess{\'e}n.\cite{essen11}

\subsection{What about Larmor's theorem?}
Larmor's theorem states that a spherically symmetric system of charged particles will start to
rotate if an external magnetic field is turned on (see Landau and Lifshitz\cite{BKlandau2} \S 45).
The rotation of such a system produces a circulating current and thus a magnetic field. Simple
calculations show that this field is of opposite direction to the external field and so the system
is diamagnetic. This idea was first used by Langevin to derive diamagnetism. But from the point
of view of the Bohr-van Leeuwen theorem this seems strange. In Feynman's lectures
\cite{BKfeynman&al2} the Bohr-van Leeuwen theorem is therefore not considered to be valid for
systems that can rotate.  As noted above, the problem is simply that the Bohr-van Leeuwen theorem
does not take into account the magnetic field produced by the particles of the system itself. When
this internal field is accounted for, diamagnetism follows naturally, whether for systems of atoms
from Larmor's theorem\cite{essen89} or for superconductors from the Darwin
formalism.\cite{essen05}

\subsection{The Shanghai experiment - measuring a diamagnetic current?}
If it is true that the classical Hamiltonian for a system of charged particles predicts
diamagnetism, then why is the phenomenon so weak and insignificant in most cases?  Is there any
evidence for classical diamagnetism for systems other than superconductors?  As stressed by
Mahajan,\cite{SCmahajan} plasmas are typically diamagnetic. However, plasmas are not usually in
thermal equilibrium so it is difficult to reach any definite conclusions from them. An experiment on an electron gas in thermal equilibrium, performed by Xinyong Fu and Zitao Fu\cite{fu&fu} in Shanghai, is therefore of considerable interest.

Two electrodes of Ag-O-Cs side-by-side in a vacuum tube emit electrons at room temperature because
of their low work function. If a magnetic field is imposed on this system, an asymmetry arises
and electrons flow from one electrode to the other. For a field strength of about 4~gauss a steady current of $\sim\!10^{-14}\,$A is measured at room temperature. The current grows with increasing field strength and changes direction as the polarity is reversed. The authors interpret this result as if the magnetic field acts as a Maxwell demon that can violate the second law of thermodynamics.\cite{fu&fu} In view of statistical mechanics based on the Darwin Hamiltonian,\cite{essen11} it is more natural to interpret this result as a diamagnetic response of the system. The current---just as the super-current of the Meissner effect---is due to a diamagnetic thermal equilibrium. This means that no useful work can  be extracted from the system.

\section{On the alleged inconsistency of magnetic flux expulsion with classical physics}
Meissner and Ochsenfeld \cite{SCmeissner,SCforrest} discovered the Meissner effect in 1933. To their
surprise a magnetic field was not only unable to penetrate a superconductor, it was also expelled
from the interior of a conductor as it was cooled below its critical temperature. The first
effect---ideal diamagnetism---seemed natural to them even though it violates the Bohr-van Leeuwen
theorem.  As already discussed, by violating this theorem ideal diamagnetism should have been considered a non-classical effect. The flux expulsion, on the other hand, was explicitly proclaimed by Meissner
to have no classical explanation. Meissner does not give any arguments or references to support this
statement, but according to Dahl\cite{BKdahl} the theoretical basis was Lippmann's theorem on the
conservation of magnetic flux through an ideally conducting current loop. Later, Forrest
\cite{SCforrest} and others have argued that the magnetohydrodynamic theorem on frozen-in flux lines also supports this notion. In this section we discuss these arguments and point out that they do not rule out a classical explanation of the Meissner effect.

\subsection{Lippmann's theorem}
Gabriel Lippmann (1845-1921), winner of the 1908 Nobel Prize in Physics,  published a theorem in
1889 stating that the magnetic flux through an ideally conducting current loop is conserved. In
1919, when superconductivity had been discovered, Lippmann \cite{SClippmann} again published this
result in three different French journals (see Sauer \cite{SCsauer}). Although the idea of flux conservation is considered highly fundamental,\cite{BKhehl&obukhov} nowadays references to Lippman's theorem are hard to find.  But at the time Lippman's theorem was quite influential, and P. F. Dahl \cite{BKdahl} explains how this was one of the results that made the Meissner effect seem surprising---and non-classical---at the time of its discovery.

The proof of Lippmann's theorem follows by noting that the self inductance $L$ of a closed circuit, or loop, is related to the magnetic flux $\Phi$ from the current in the loop via
\begin{equation}\label{eq.flux.self.induct}
\Phi = c L \dot q ,
\end{equation}
where $c$ is the speed of light and $\dot q$ is the current through the circuit, an over dot denoting a time derivative (see Landau and Lifshitz\cite{BKlandau8} vol.\ 8 \S 33). The equation of motion for a single loop electric circuit is
\begin{equation}\label{eq.circuit}
L \ddot q + R \dot q + C^{-1} q = {\cal E}(t),
\end{equation}
where $R$ is the resistance, $C$ the capacitance, and ${\cal E}(t)$ the emf driving the current.
If there is no resistance, no capacitance, and no emf, this equation becomes
\begin{equation}\label{eq.circuit.null}
L \ddot q = 0.
\end{equation}
Therefore, if the self inductance $L$ is constant, Eqs.~(\ref{eq.flux.self.induct}) and (\ref{eq.circuit.null}) tell us that $\Phi =\, $constant.

\subsection{Lippmann's theorem and superconductors}
Although Lippmann's theorem is correct, its relevance for the prevention of flux expulsion
is not clear. For superconductors there are two points to consider, the assumption of zero emf, and the fact that constant flux does not imply constant magnetic field.  We consider these points one at a time.

Consider a superconducting sphere of radius $r$ in a constant external magnetic field $\vecB_{\rm e}$. When the sphere expels this field by generating (surface) currents that produce $\vecB_{\rm i}=-\vecB_{\rm e}$ in its interior, the total magnetic energy is reduced. The
magnetic energy change is
\begin{equation}\label{eq.energy.reduction.in.sphere}
\Delta E_B = -3\left(\frac{4\pi r^3}{3}\right) \frac{\vecB_{\rm e}^2}{8\pi},
\end{equation}
or three times the initial interior magnetic energy.\cite{SCfiolhais&al} This energy is thus
available for producing the emf required to generate currents in the sphere's interior. The
assumption of Lippmann's theorem---that the emf is zero---is therefore not fulfilled.

When a steady current flows through a fixed metal wire both the flux and the magnetic field distribution are constant. On the other hand, when current flows in a loop in a conducting medium, a constant flux through the loop does not imply a constant magnetic field because the loop can change in size, shape, or location. Normally current loops are subject to forces that increase their radius (see \eg\ Landau and Lifshitz\cite{BKlandau8} vol.\ 8 \S 34 Problem 4, or Ess{\'e}n \cite{essen09} Section 4.1). In view of this, Lippmann's theorem does not automatically imply that the magnetic field must be constant, even if the emf is zero.

Other authors have reached similar conclusions. Mei and Liang \cite{SCmei&liang} carefully
considered the electromagnetics of superconductors in 1991 and write, ``Thus Meissner's experiment
should be viewed through its time history instead of as a strictly dc event. In that case classical
electromagnetic theory will be consistent with the Meissner effect.''

\subsection{Frozen-in field lines}
The simplest derivation of the frozen-in field result for a conducting medium begins with Ohm's law $\vecj = \sigma \vecE$.  If $\sigma \rightarrow \infty$ we must have $\vecE = \vecnl$ to prevent infinite current. Faraday's law then gives $\partial\vecB/\partial t =-c(\nabla\times\vecE)=\vecnl$ which tells us the magnetic field $\vecB$ is constant (Forrest,\cite{SCforrest} Alfv{\'e}n and F{\"a}lthammar\cite{BKalfven}).

When dealing with a conducting fluid, the equations of magnetohydrodynamics and the limit of infinite conductivity in Ohm's law gives rise to\cite{BKalfven}
\begin{equation}
\label{eq.frozen.in.field.lines}
\frac{\partial \vecB}{\partial t} = \nabla\times(\vecv\times\vecB).
\end{equation}
This result tells us the magnetic field convects with the fluid but does not dissipate.  In addition, Eq.~(\ref{eq.frozen.in.field.lines}) can be used to derive the following two statements that are often referred to as the frozen-in field theorem:\cite{BKgurnett}
\begin{enumerate}
\item the magnetic flux through any closed curve moving with the fluid is constant, and
\item a magnetic field line moving with the fluid remains a magnetic field line for all time.
\end{enumerate}
The first of these is a just a restatement of Lippmann's theorem for a current loop in a fluid.

\subsection{Magnetic field lines in superconductors}
As shown in the previous section, Ohm's law with infinite conductivity implies there can be no change in the magnetic field since this would give rise to an infinite current density. It is, however, not physically meaningful to take the limit $\sigma \rightarrow \infty$ in Ohm's law. In a medium of zero resistivity one must instead use the equation of motion for the charge carriers. One can then derive\cite{SCfiolhais&al,SCbadiamajos}
\begin{equation}
\label{eq.time.deriv.j}
\frac{\dfd \vecj}{\dfd\, t} = \frac{e^2 n}{m} \vecE +
\frac{e}{mc} \vecj\times\vecB ,
\end{equation}
for the time rate of change of the current density. The time derivative here is a convective, or material, time derivative. Ohm's law simply does not apply, and therefore it cannot imply that the magnetic field does not change.

The second statement of the frozen-in field theorem can be questioned on the same grounds. Just as with Lippmann's theorem, the conclusion that the magnetic field remains constant does not follow, even if
the statement is assumed valid. The magnetic field is determined by the density of magnetic field lines and constancy of this density requires that the conducting fluid is incompressible. One can easily imagine that the fluid of superconducting electrons is compressible, and that magnetic pressure \cite{BKgurnett}
pushes the fluid (with its magnetic field lines) to the surface of the material. Corroborating this point of view, Alfv\'en and F\"althammar \cite{BKalfven} (section 5.4.2) state that ``in low density plasmas
the concept of frozen-in lines of force is questionable.''

\subsection{A classical derivation of the London equations}
In 1981 W.\ Farrell Edwards\cite{SCedwards} published a manuscript with the above title in Physical Review Letters. This caused an uproar of indignation and the journal
later published three different criticisms of Edwards' work.\cite{SChenyey,SCsegal&al,SCdevegvar} Incidentally, all of Edwards' critics restated (or implied) the textbook myth that the Meissner flux expulsion does not have a classical explanation.  In addition, the journal Nature published a study by John Bryan Taylor \cite{SCtaylor} pointing out the faults of Edwards' derivation. Of course, Edwards is not the only one to publish an erroneous derivation of the London equations. A much earlier example is the derivation by Moore\cite{SCmoore} from 1976.  The fact that various derivations have been wrong, of course, does not prove anything---as long as a correct derivation exists.

In 1966 Nature published a classical explanation of the London equations by
Pfleiderer \cite{SCpfleiderer} which did not cause any comment. Indeed it has not been cited a
single time, probably because it is very brief and cryptic. However, a classical derivation of the London equations and flux expulsion can be found in a classic textbook by the French Nobel laureate Pierre Gilles de Gennes.\cite{BKdegennes} Because this derivation should be more recognized, we repeat it here.

\subsection{de Gennes' derivation of flux expulsion}
The total energy of the relevant electrons in the superconductor is assumed to have three
contributions: the condensation energy associated with the phase transition $E_S$, the energy of the
magnetic field $E_B$, and the kinetic energy of the moving superconducting electrons $E_k$. The
total energy relevant to the problem is thus taken to be
\begin{equation}\label{eq.energy.SC.tot}
E = E_S + E_B + E_k .
\end{equation}
The condensation energy is then assumed to be constant while the remaining two can vary in response
to external field variations. The super-current density is written as
\begin{equation}\label{eq.sup.cond.curr}
\vecj(\vecr) = n(\vecr)\, e\, \vecv(\vecr),
\end{equation}
where $n$ is the number density of superconducting electrons
and $\vecv$ is their velocity, which gives a kinetic energy of
\begin{equation}\label{eq.kin.energy}
E_k = \int \half n(\vecr) m \vecv^2(\vecr)\,\dfd V = \int \half \frac{m}{e^2\, n(\vecr)}
\vecj^2(\vecr)\,\dfd V .
\end{equation}
By means of the Maxwell equation $\nabla\times\vecB = 4\pi\vecj /c$ and Eq.~(\ref{eq.mag.energy.1.2}) for $E_B$, the total energy (\ref{eq.energy.SC.tot}) becomes
%\begin{equation}\label{eq.maxwell.eq}
%\nabla\times\vecB = \frac{4\pi}{c} \vecj,
%\end{equation}
\begin{equation}\label{eq.energy.SC.tot.expl}
E = E_S + \frac{1}{8\pi} \int \left[\vecB^2 + \lambda^2 (\nabla\times\vecB)^2 \right]\, \dfd V,
\end{equation}
where we have assumed that $n$ is constant in the region where there is current, and the London penetration depth is given by
\begin{equation}\label{eq.lambda.sq.def}
\lambda = \sqrt{\frac{mc^2}{4\pi\,e^2 n }}.
\end{equation}
Minimizing the energy in Eq.~(\ref{eq.energy.SC.tot.expl}) with respect to $\vecB$ then gives the London equation
\begin{equation}\label{eq.London.according.to.degennes}
\vecB + \lambda^2\, \nabla\times(\nabla\times\vecB) = \vecnl,
\end{equation}
in one of its equivalent forms. Notice that this derivation utilizes no
quantum concepts and does not contain Planck's constant. It is thus completely classical. A similar
derivation has been published more recently by Bad{\'i}a-Maj{\'o}s.\cite{SCbadiamajos}

The conclusion of de Gennes is clearly stated in his 1965 book\cite{BKdegennes}  (emphasis from the original): {\em ``The superconductor finds an equilibrium state where the sum of the kinetic and magnetic energies is minimum, and this state, for macroscopic samples, corresponds to the expulsion of magnetic flux.''}  In spite of this, most textbooks continue to state that ``flux expulsion has no classical explanation'' as originally stated by Meissner and Ochsenfeld \cite{SCmeissner} and repeated in the influential monographs by Fritz London \cite{BKlondon} and Nobel laureate Max von Laue.\cite{BKvonlaue}  As one textbook example, Ashcroft and Mermin\cite{BKashcroft&mermin} explain that ``perfect conductivity implies a time-independent magnetic field in the interior.''

\subsection{A purely classical derivation from magnetostatics}
One might object that the electronic charge $e$ is a microscopic constant, and that the London
penetration depth $\lambda$ in most cases is so small that it seems to belong to the domain of
microphysics. So even if quantum concepts do not enter explicitly, the above derivation does have
microscopic elements. It is thus of interest that from a purely macroscopic point of view we can
identify the kinetic energy of the conduction electrons solely with magnetic
energy.\cite{darwin2,SCcullwick2} The energy that should be minimized would then be
Eq.~(\ref{eq.mag.energy.1.2})
\begin{equation}\label{eq.mag.energy.1.3}
E_B = \frac{1}{8\pi} \int \vecB^2\, \dfd V.
\end{equation}
Unfortunately, this will not provide any information about the currents that are the sources of
$\vecB$. However, if we can neglect the contribution from fields at a sufficiently distant
surface---\ie\ when radiation is negligible---it is possible to rewrite this expression as
\begin{equation}\label{eq.mag.energy.1.4}
E'_B = \frac{1}{2c} \int \vecj\cdot\vecA\, \dfd V,
\end{equation}
where $\vecB=\nabla\times\vecA$. The idea is to then apply the variational principle to $E_{jA}=2E'_B
- E_B$, \ie\ the expression
\begin{equation}\label{eq.mag.energy.1.5}
E_{jA} =  \int \left[\frac{1}{c}\vecj\cdot\vecA -\frac{1}{8\pi}\left(\nabla\times\vecA\right)^2
\right] \dfd V .
\end{equation}
Here the energy is written in terms of the field $\vecA$ and its source $\vecj$. There are many ways of combining (\ref{eq.mag.energy.1.3}) and (\ref{eq.mag.energy.1.4}) to get an energy expression, but it turns out that (\ref{eq.mag.energy.1.5}) is the one that gives the simplest results.

Starting from (\ref{eq.mag.energy.1.5}) and adding the constraint $\nabla\cdot\vecj =0$, the integration is split into interior, surface, and exterior regions. The result is a theorem of classical
magnetostatics that states, {\em for a system of perfect conductors the magnetic field is zero
in the interiors and all current flows on their surfaces} (Fiolhais et al.\ \cite{SCfiolhais&al})
This theorem is analogous to a similar result in electrostatics for electric fields and charge
densities in conductors, sometimes referred to as Thomson's theorem (see Stratton,\cite{BKstratton}
Sec.\ 2.11).  Readers should consult the paper by Fiolhais et al.\ \cite{SCfiolhais&al} for details.

Many other authors have also reached the conclusion that the equilibrium state of a superconductor is in fact the state of minimum magnetic energy of an ideal conductor. Some of these are, in chronological order, Cullwick,\cite{SCcullwick} Pfleiderer,\cite{SCpfleiderer} Karlsson,\cite{SCkarlsson} Bad{\'i}a-Maj{\'o}s, \cite{SCbadiamajos} Kudinov, \cite{SCkudinov} and Mahajan.\cite{SCmahajan} For example, Cullwick writes ``The well-known Meissner effect in pure superconductors is shown to be an expected rather than an unexpected phenomenon...'' while Kudinov\cite{SCkudinov} explains that it ``is worth noting in this  connection that the expulsion of a magnetic field to the periphery (the Meissner effect) also occurs in a classical collisionless gas of charged particles and that this happens solely because this state is energetically favorable.''

\section{Conclusions}
The reader may get the impression from our investigations above that we consider superconductivity to be a classical phenomenon. Nothing could be further from the truth. As implied by the Ginzburg-Landau theory,\cite{SCginzburg&landau} the BCS theory,\cite{SCbcs} and the Josephson effect,\cite{SCjosephson} the phenomenon is quantum mechanical to a large extent. Since quantum physics must lead to classical physics in some macroscopic limit, it must be possible to derive our classical result from a quantum perspective. Evans and Rickayzen\cite{SCevans&rickayzen} did indeed derive the equivalence of zero resistivity and the Meissner effect quantum mechanically, but did not discuss the classical limit. What we want to correct is the mis-statement that the Meissner effect proves that superconductors are ``not just perfect conductors.'' According to basic physics and a large number of independent investigators, the specific phenomenon of flux expulsion follows naturally from classical physics and the zero resistance property of the superconductor---they \textit {are} just perfect conductors.

In conclusion, we have carefully examined the evidence for the oft repeated statements in textbooks
that 1) there is no classical diamagnetism and 2) there is no classical explanation of magnetic flux
expulsion. We have found that the theoretical arguments for these statements are not rigorous and
that, for superconductors, these particular phenomena are in good agreement with the classical
physics of ideal conductors.

%\bibliography{ClassMeissner,BohrVanLeeuwenBib,Books,DarwinPapers}

\begin{thebibliography}{10}
\bibitem{Footnote12} We only consider
magnetism from moving charged particles, not from dipole densities. We therefore need not
distinguish between the $\vecB$ and the $\vecH$-fields. Gaussian units are used.
\end{thebibliography}

\end{document}